# Growth of $(Na_xK_y)Fe_zSe_2$ crystals by chlorides flux at low temperatures

*Gang Wang[a], Tianping Ying[a], Yaobo Huang[b], Shifeng Jin[a], Lei Yan[b], Hong Ding[b,c], Xiaolong Chen[a,c,] \**

[a] Research & Development Center for Functional Crystals, Beijing National Laboratory for Condensed Matter Physics, Institute of Physics, Chinese Academy of Sciences, Beijing 100190, China

[b] Key Laboratory of Extreme Conditions Physics, Beijing National Laboratory for Condensed Matter Physics, Institute of Physics, Chinese Academy of Sciences, Beijing 100190, China

[c] Collaborative Innovation Center of Quantum Matter, Beijing, China

*Corresponding Author: No.8 Nansanjie, Zhongguancun, Haidian District, Beijing 100190, China. Tel: + 86 10 8264 9039; Fax: + 86 10 8264 9646.

E-mail address: chenx29@iphy.ac.cn

**ABSTRACT:** $(Na_xK_y)Fe_zSe_2$ crystals are prepared by Na, Fe, and Se as starting materials in NaCl/KCl flux at low temperatures~720 °C. It is found that K is more preferred than Na to enter in between FeSe layers and forms the phase. Thus-obtained crystals contain more superconducting phase in volume fraction and exhibit new features in transport property. Our results provide a promising new synthetic route for preparing quality crystals of iron selenide superconductors.

## 1. Introduction

Since the discovery of high temperature superconductivity in cuprate [1], many new families of layered superconductors have been discovered [2-4]. The very new member of these families is metal-intercalated iron selenides, like $K_xFe_2Se_2$ [5], $Cs_{0.8}(FeSe_{0.98})_2$ [6], $Rb_xFe_ySe_2$ [7], $Tl_{0.58}Rb_{0.42}Fe_{1.72}Se_2$ [8], and $(Tl,K)Fe_xSe_2$ [9]. Compared to their iron pnictide counterparts, these superconductors display several remarkably distinct structural and physical characteristics. It was demonstrated by angle-resolved photoemission spectroscopy (ARPES) that there are only electron Fermi surfaces observed at the Brillouin zone corners while the hole bands at the center of the Brillouin zone sink below the Fermi level ($E_F$) [10-13], unlike other iron-based superconductors with both electron and hole Fermi surfaces locating at the Brillouin zone corners and center, respectively [14,15]. Single crystal X-ray diffraction [16] and scanning tunneling microscope [17] measurements indicated that the main phases have $\sqrt{5} \times \sqrt{5} \times 1$ superstructures (insulating phases). Neutron diffraction further revealed that the Fe vacancies order to form supercells above 500 K and then antiferromagnetic order at a little bit lower temperatures with large magnetic moments of about 3.3 $\mu_B$/Fe at 10 K [18,19]. However, the presence of phase separations in the materials evidenced by various techniques [20-24] severely thwarts the attempts to unambiguously identify the superconducting (SC) phases and to understand the underlying mechanisms. Moreover, the occasionally observed 44 K SC transition further complicated the issue [25]. Many efforts have been devoted to obtain phase-pure or SC-dominated single crystals, such as self-flux method [25-27] and optical floating zone technique [28]. Recently, we succeeded in synthesizing a series of superconductors $A_xFe_2Se_2(NH_3)_y$ (A = Li, Na, K, Ba, Sr, Ca, Eu, and Yb) with transition temperatures ($T_c$s) of 30~46 K and clarifying the SC phases in K-Fe-Se system via a liquid ammonia route [29,30], which imply that low temperature is an alternative for preventing the phase separation. The shortage of this route, however, is difficult to obtain single crystal of SC phase. Here we report a new approach to grow iron selenide superconductors by flux method at relatively low temperature of 720 °C, much lower than the temperatures (~1040 °C) by high-temperature melting method. The flux consisting of NaCl and KCl effectively lowers the melting point of the system. Thus-obtained crystals contain both Na and K, though K dominates, and exhibit new features in transport property for higher volume fraction of SC phase.

## 2. Experimental section

Previously, NaCl/KCl, KCl, LiCl/CsCl, and KCl/AlCl$_3$ were used as flux to grow FeSe single crystals [31-36]. In this study, we choose NaCl/KCl as flux and elements Na, Fe, and Se as the starting materials. Na piece, Fe powder, Se powder, NaCl and KCl powders with predetermined molar ratios were loaded into alumina crucibles. The NaCl/KCl/Na$_m$Fe$_n$Se$_2$ ratio was 10:10:1. NaCl and KCl powders were heated to 130 °C and kept 24 h prior to loading into the alumina crucibles to eliminate the moisture. The alumina crucibles were sealed within the evacuated quartz ampoules partially backfilled with high-purity argon. The ampoules were loaded into a Muffle furnace and slowly raised to 720 °C and kept for 30 h. Then the ampoules were cooled at 0.5 °C/h from 720 °C to 650 °C and then held at 650 °C for 50 h. At last, they were furnace cooled to room temperature. The ampoules were broken in glove box and crystals embedded in the solution were separated. The crystal morphology and composition were characterized by optical microscopy, scanning electron microscopy (SEM, Hitachi S-4800) equipped with an energy dispersive X-ray (EDX) spectrometer, and inductively coupled plasma-atomic emission spectroscopy (ICP-AES). X-ray diffraction (XRD) was performed at room temperature using a PANalytical X'Pert PRO diffractometer equipped with a graphite monochromator utilizing Cu $K\alpha$ radiation (40 kV, 40 mA). Rietveld refinements were performed with the FULLPROF package [37]. The magnetic properties of accurately weighed sample in a gelatin capsule were characterized in DC field of 40 Oe in the temperature ranging from 5 to 300 K using a vibrating sample magnetometer (VSM, Quantum Design) which features a sensitivity of $10^{-7}$ emu. The electrical resistance was measured by four probe method using the physical property measurement system (PPMS, Quantum Design). The ARPES measurements were performed using a VG-Scienta R4000 electron analyzer with the He $I\alpha$ ($h\nu$ = 21.2 eV) resonance line. The angular resolution was set to 0.2° and the energy resolution to 10 and 25 meV for band structure and Fermi surface mapping respectively. The sample was cleaved in-situ and measured at 10 K under vacuum better than $5\times10^{-11}$ torr.

## 3. Results and discussion

The inset to Figure 1 shows the optical image of the as-prepared crystal embedded in the solution. The crystal exhibits a plate-like morphology with a shining surface. The compositions of washed samples grown from different nominal compositions are characterized by ICP-AES and listed in Table 1. The measurements reveal that K has a much higher concentration in crystals than Na. Figure 2(a) shows the typical SEM image of $(Na_{0.16}K_{0.70})Fe_{1.72}Se_2$ crystal with SC stripes embedded in insulating background. The EDX results (Figure 2(b) and (c)) reveal that the Na/K ratio and iron content are relatively higher in SC stripes while the total amount of (Na+K) is less than that of the background. From Figure 2(c), we can see the Na's concentration in SC stripes is about 1% higher than that of insulating background for these randomly selected points. We performed the EDX mapping of Na on SC stripes and insulating background, but the data are quite scattered. The reason is not clear. The XRD pattern of $(Na_{0.16}K_{0.70})Fe_{1.72}Se_2$ crystal, as shown in Figure 1, is dominated by the $(00\ell)$ ($\ell = 2n$) reflections, revealing the surface of the crystal is perpendicular to crystallographic $c$ axis. The reflections marked by asterisk are assigned to the SC phase as its lattice constant $c$ is a little bit larger than that of the insulating phase [24]. The powder XRD pattern (Figure 1S) of crushed crystals with composition of $(Na_{0.18}K_{0.66})Fe_{1.68}Se_2$ was refined based on a multi-phase scenario, i.e., SC phase coexisting with insulating phase and a tiny fraction of flux by using the Rietveld method. Only the structural parameters of SC phase are refined while the insulating phase's and flux's parameters are fixed as their structures are well known. The refined lattice parameters of SC phase are $a = 3.8924(6)$ Å, $c = 14.111(3)$ Å and $V = 214.80(6)$ Å$^3$. All crystallographic data are summarized in Table 1S.

A surprising result is that the product crystals are K rich and Na poor in composition though no metal K in the starting materials. We presume that KCl is reduced to metallic K through reaction (1) at temperatures not higher than 720 °C. The metallic K then enters in between FeSe layers and finally crystallizes into $(Na_xK_y)Fe_zSe_2$ crystals.

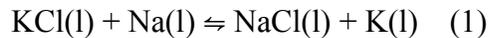

$$KCl(l) + Na(l) \leftrightharpoons NaCl(l) + K(l) \quad (1)$$

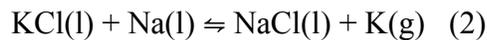

$$KCl(l) + Na(l) \leftrightharpoons NaCl(l) + K(g) \quad (2)$$

It is known that only at above 850 °C, K ion can be reduced to metallic K by Na through reaction (2). This is due to that liquid K gasifies beyond its boiling point and leaves the system, tipping the equilibrium of reaction (2) and makes the reduction reaction to complete. In the present case, the driven force of reaction (1) comes from liquid K consuming by its intercalation in between FeSe layers, which is an energy-favored process. On the other hand, Na is found to be difficult to insert into FeSe layers and form the compound with a body-centered tetragonal structure at relative high temperatures due to the instability of lattice dynamics based on our experimental results and first-principles calculations. This explains why K is much more preferred than Na to enter into FeSe layers. An alternative interpretation is that elemental Na, Se, and Fe are likely to react first and form metastable Na-Fe-Se ternary compounds. Then the ternary compounds dissolve in the melting chlorides and $(Na_xK_y)Fe_zSe_2$ crystals finally crystallize upon cooling. We tried K as the starting material instead of Na, but with no discernible crystals identified in products. At present, the reason is not clear.

We examined the superconductivity in the obtained crystals. The magnetization of $(Na_{0.16}K_{0.70})Fe_{1.72}Se_2$ with temperature ranging from 5 to 80 K is shown in Figure 3. The external applied magnetic field was set perpendicular to the plane of the crystal. The zero field cooling (ZFC) magnetization vs temperature curve reveals that $(Na_{0.16}K_{0.70})Fe_{1.72}Se_2$ grown from NaCl/KCl flux exhibits superconductivity with $T_c^{onset} = 29$ K. The estimated SC shielding fraction is about 62% at 10 K. The *M-H* curve at 5 K (inset of Figure 3) indicates a typical feature of Type-II superconductor. The lower critical magnetic field ($H_{c1}$) is estimated to be around 0.2 T. In-plane electrical resistance vs temperature plot of $(Na_{0.16}K_{0.70})Fe_{1.72}Se_2$ is shown in Figure 4, further confirming the SC phase coexists with the insulating phase. The curve first exhibits semiconducting behaviour at temperatures above 212 K and a large hump centered at around this temperature ($T_H$), which is twice as high as that (105 K) of $K_{0.8}Fe_2Se_2$ [5]. Below $T_H$, the curve displaying metallic behaviour shows a SC transition at 29.5 K and reaches zero resistance at 27.4 K (shown in the inset). The substitution of cation has little effect on the SC transition temperature, in contrast to $(Ba_{1-x}K_x)Fe_2As_2$ and so on [38,39]. Following the model proposed by Shoemaker *et al.* [24], we fitted the resistance above $T_c^{onset}$ against temperature very well using equation (3). $n = 2.238$ and $E_g = 107$ meV are obtained.

$$\frac{1}{R_{total}} = \frac{1}{R_1} + \frac{1}{R_2} = \frac{1}{[R(0) + AT^n]} + R_0^{-1} e^{-\frac{E_g}{2kT}} \quad (3)$$

Where *R(0)*, *A*, and *R₀* are constants that depend on phase fractions and geometry of SC phase and insulating phase. The $T_H$ more than two times higher than 105 K can be interpreted by an increase of volume fraction of the SC phase, which is consistent with the estimated superconductive shielding fraction. This enhanced SC phase/insulating phase ratio can be attributed to the lower growth temperature that is helpful to restrain the formation of insulating phase to some extent.

Figure 5 shows the photoemission intensity maps at the Fermi level ($E_F$) and the band dispersions along $\Gamma$-$M$ for $(Na_{0.16}K_{0.70})Fe_{1.72}Se_2$. As shown in Figure 5(a), the Fermi surface which is defined by the high intensity region includes a nearly circular electron-like pocket centered at each Brillouin zone corner $M$ point. There is no Fermi surface pocket opening found at Brillouin zone center $\Gamma$ but some spectral weight that comes from the tail of the electron-like band bottom which sits closely above $E_F$ at the $k_z$ where the detecting photon energy 21.2 eV corresponds to. This upper electron-like band was predicted by the local-density-approximation (LDA) calculation on the similar material $(K,Cs)_xFe_2Se_2$ [40], and with a significant $k_z$ dispersion it opens as a pocket at $k_z = \pi$. This $k_z$ dispersion has been proved in the studies of $K_{0.8}Fe_2Se_2$ [11] and $(Tl,K)Fe_{1.78}Se_2$ [13] by the photon energy dependent ARPES, where one maps the band dispersion along $k_z$ by tuning the photon energy. The diameter of the circular electron-like Fermi surface pocket is about $0.552\pi/a$, corresponding to 6.0% of the Brillouin zone area. From the LDA calculation on $(K,Cs)_xFe_2Se_2$ [40] and all the previous ARPES measurements on other iron-based superconductors [41], two nearly degenerate electron-like pockets centered at $M$ are suggested. Regarding the crystal structure of $(Na_{0.16}K_{0.70})Fe_{1.72}Se_2$ is quite similar as that of $(K,Cs)_xFe_2Se_2$ [40], it is reasonable to assume that the observed electron-like pockets are actually two which overlap with each other due to the degeneracy. This assumption also have been proposed in previous ARPES on $K_{0.8}Fe_2Se_2$ [10,11]. Thus the overall electron concentration is estimated to be about 0.12 per Fe atom, if ignores the contribution from the trace of the electron-like band weight at $\Gamma$, which is reasonable from the point of optimal doping view.

The band dispersions are further revealed in Figure 5(b)-(e), where (b)/(d) shows the intensity plot for the cuts passing through the Brillouin zone center-$\Gamma$/corner-$M$ along $\Gamma$-$M$ direction, respectively, and (c)/(e) displays the corresponding momentum-distribution-curves (MDCs) to (b)/(d). In Figure 5(b), close to $E_F$, a bit spectral weight coming from the upper electron-like band named $\kappa$ as discussed above is observed at $\Gamma$. Meanwhile, a hole-like band named $\alpha$ with band top at around 80 meV below $E_F$ is also found. These band features are more clearly to see in the MDCs in Figure 5(c). In contrast to the $\Gamma$ point, an electron-like band is clearly seen crossing $E_F$ passing through the $M$ point with its bottom sitting at about 55 meV below $E_F$ as shown in the intensity plot in Figure 5(d) and the MDCs in Figure 5(e). It should be pointed out that despite we examine the Fermi surface and the band dispersions very carefully, no folded band structures corresponding to the $\sqrt{5}\times\sqrt{5}\times1$ superstructure, which is claimed to be related to the insulating phase [42], is found in the recorded energy within 200 meV below the $E_F$. It is reported by the ARPES study on $K_xFe_{2-y}Se_2$ [42], with a systematic review on the band structure of different phases at different temperatures, the valence band of the insulating phase is sitting far away (> 500 meV) below the $E_F$ and mainly contributes to the higher banding-energy density-of-states (DOS), while the SC phase mainly contribute to the lower energy DOS from 400 meV to $E_F$. From the similar band structure point of view, we believe that the observed band dispersions in this study only represents the low energy band structure of the SC phase for the as grown $(Na_{0.16}K_{0.70})Fe_{1.72}Se_2$ sample thus does not show the $\sqrt{5}\times\sqrt{5}\times1$ order.

## 4. Conclusions

In conclusion, we prepare $(Na_xK_y)Fe_zSe_2$ crystals with different compositions using NaCl/KCl flux at relatively low temperatures. The obtained crystals contain more SC phase in volume fraction and exhibit new features in transport property. These results imply the potential of flux method as an effective route for the preparation of single crystals of iron selenide superconductors.

## ACKNOWLEDGMENT

G. Wang would like to thank W. H. Zhang of Tsinghua University, S. C. Wang, J. Shen, and Z. L. Li of the Institute of Physics, Chinese Academy of Sciences for assistance in sample preparation, resistance measurements and data fitting. This work was partly supported by the National Natural Science Foundation of China under Grant Nos. 51322211, 51272276, 51072222, and 90922037, Beijing Nova Program (2011096), and the Chinese Academy of Sciences.

## REFERENCES

(1) Bednorz, J. G.; Muller, K. A. *Z. Phys. B: Condens. Matter* **1986**, *64*, 189-193.

(2) Nagamatsu, J.; Nakagawa, N.; Muranaka, T.; Zenitani, Y.; Akimitsu, J. *Nature* **2001**, *410*, 63-64.

(3) Takada, K.; Sakurai, H.; Takayama-Muromachi, E.; Izumi, F.; Dilanian, R. A.; Sasaki, T. *Nature* **2003**, *422*, 53-55.

(4) Kamihara, Y.; Watanabe, T.; Hirano, M.; Hosono, H. *J. Am. Chem. Soc.* **2008**, *130*, 3296-3297.

(5) Guo, J. G.; Jin, S. F.; Wang, G.; Wang, S. C.; Zhu, K. X.; Zhou, T. T.; He, M.; Chen, X. L. *Phys. Rev. B* **2010**, *82*, 180520(R).

(6) Krzton-Maziopa, A.; Shermadini, Z.; Pomjakushina, E.; Pomjakushin, V.; Bendele, M.; Amato, A.; Khasanov, R.; Luetkens, H.; Conder, K. *J. Phys.: Condens. Matter* **2011**, *23*, 052203.

(7) Wang, A. F.; Ying, J. J.; Yan, Y. J.; Liu, R. H.; Luo, X. G.; Li, Z. Y.; Wang, X. F.; Zhang, M.; Ye, G. J.; Cheng, P.; Xiang, Z. J.; Chen, X. H. *Phys. Rev. B* **2011**, *83*, 060512.

(8) Wang, H.-D.; Dong, C.-H.; Li, Z.-J.; Mao, Q.-H.; Zhu, S.-S.; Feng, C.-M.; Yuan, H. Q.; Fang, M.-H. *Europhys. Lett.* **2011**, *93*, 47004.

(9) Fang, M.-H.; Wang, H.-D.; Dong, C.-H.; Li, Z.-J.; Feng, C.-M.; Chen, J.; Yuan, H. Q. *Europhys. Lett.* **2011**, *94*, 27009.

(10) Qian, T.; Wang, X.-P.; Jin, W.-C.; Zhang, P.; Richard, P.; Xu, G.; Dai, X.; Fang, Z.; Guo, J.-G.; Chen, X.-L.; Ding, H. *Phys. Rev. Lett.* **2011**, *106*. 187001.

(11) Zhang, Y.; Yang, L. X.; Xu, M.; Ye, Z. R.; Chen, F.; He, C.; Xu, H. C.; Jiang, J.; Xie, B. P.; Ying, J. J.; Wang, X. F.; Chen, X. H.; Hu, J. P.; Matsunami, M.; Kimura, S.; Feng, D. L. *Nat. Mater.* **2011**, *10*, 273-277.

(12) Zhao, L.; Mou, D. X.; Liu, S. Y.; Jia, X. W.; He, J. F.; Peng, Y. Y.; Yu, L.; Liu, X.; Liu, G. D.; He, S.; Dong, X. L.; Zhang, J.; He, J. B.; Wang, D. M.; Chen, G. F.; Guo, J. G.; Chen, X. L.; Wang, X. Y.; Peng, Q. J.; Wang, Z. M.; Zhang, S. J.; Yang, F.; Xu, Z. Y.; Chen, C. T.; Zhou, X. J. *Phys. Rev. B* **2011**, *83*, 140508.

(13) Wang, X.-P.; Richard, P.; Shi, X.; Roekeghem, A.; Huang, Y.-B.; Razzoli, E.; Qian, T.; Rienks, E.; Thirupathaiah, S.; Wang, H.-D.; Dong, C.-H.; Fang, M.-H.; Shi, M.; Ding, H. *Europhys. Lett.* **2012**, *99*, 67001.

(14) Ding, H.; Richard, P.; Nakayama, K.; Sugawara, K.; Arakane, T.; Sekiba, Y.; Takayama, A.; Souma, S.; Sato, T.; Takahashi, T.; Wang, Z.; Dai, X.; Fang, Z.; Chen, G. F.; Luo, J. L.; Wang, N. L. *Europhys. Lett.* **2008**, *83*, 47001.

(15) Terashima, K.; Sekiba, Y.; Bowen, J. H.; Nakayama, K.; Kawahara, T.; Sato, T.; Richard, P.; Xu, Y. M.; Li, L. J.; Cao, G. H.; Xu, Z.-A.; Ding, H.; Takahashi, T. *Proc. Natl. Acad. Sci* **2009**, *106*, 7330-7333.

(16) Zavalij, P.; Bao, W.; Wang, X. F.; Ying, J. J.; Chen, X. H.; Wang, D. M.; He, J. B.; Wang, X. Q.; Chen, G. F.; Hsieh, P.-Y.; Huang, Q.; Green, M. A. *Phys. Rev. B* **2011**, *83*, 132509.

(17) Li, W.; Ding, H.; Deng, P.; Chang, K.; Song, C. L.; He, K.; Wang, L. L.; Ma, X.; Hu, J.-P.; Chen, X.; Xue, Q.-K. *Nat. Phys.* **2012**, *8*, 126-130.

(18) Bao, W.; Huang, Q.-Z.; Chen, G.-F.; Green, M. A.; Wang, D.-M.; He, J.-B.; Qiu, Y.-M. *Chin. Phys. Lett.* **2011**, *28*, 086104.

(19) Ye, F.; Chi, S.; Bao, W.; Wang, X. F.; Ying, J. J.; Chen, X. H.; Wang, H. D.; Dong, C. H.; Fang, M. H. *Phys. Rev. Lett.* **2011**, *107*, 137003.

(20) Wang, Z.; Song, Y. J.; Shi, H. L.; Wang, Z. W.; Chen, Z.; Tian, H. F.; Chen, G. F.; Guo, J. G.; Yang, H. X.; Li, J. Q. *Phys. Rev. B* **2011**, *83*, 140505.

(21) Ksenofontov, V.; Wortmann, G.; Medvedev, S. A.; Tsurkan, V.; Deisenhofer, J.; Loidl, A.; Felser, C. *Phys. Rev. B* **2011**, *84*, 180508.

(22) Shermadini, Z.; Krzton-Maziopa, A.; Bendele, M.; Khasanov, R.; Luetkens, H.; Conder, K.; Pomjakushina, E.; Weyeneth, S.; Pomjakushin, V.; Bossen, O.; Amato, A. *Phys. Rev. Lett.* **2011**, *106*, 117602.

(23) Texier, Y.; Deisenhofer, J.; Tsurkan, V.; Loidl, A.; Inosov, D. S.; Friemel, G.; Bobroff, J. *Phys. Rev. Lett.* **2012**, *108*, 237002.

(24) Shoemaker D. P.; Chung D. Y.; Claus H.; Francisco M. C.; Avci S.; Llobet A.; Kanatzidis M. G. *Phys. Rev. B* **2012**, *86*, 184511.

(25) Wang, D. M.; He, J. B.; Xia, T.-L.; Chen, G. F. *Phys. Rev. B* **2011**, *83*, 132502.

(26) Ying, J. J.; Wang, X. F.; Luo, X. G.; Wang, A. F.; Zhang, M.; Yan, Y. J.; Xiang, Z. J.; Liu, R. H.; Cheng, P.; Ye, G. J.; Chen, X. H. *Phys. Rev. B* **2011**, *83*, 212502.

(27) Roslova, M.; Kuzmichev, S.; Kuzmicheva, T.; Ovchenkov, Y.; Liu, M.; Morozov, I.; Boltalin, A.; Shevelkov, A.; Chareev, D.; Vasiliev. A. CrystEngComm. **2014**, *16*, 6919–6928.

(28) Liu, Y.; Li, Z. C.; Liu, W. P.; Friemel, G.; Inosov, D. S.; Dinnebier, R. E.; Li, Z. J.; Lin. C.T. *Supercond. Sci. Technol.* **2012**, *25*, 075001.

(29) Ying, T. P.; Chen, X. L.; Wang, G.; Jin, S. F.; Zhou, T. T.; Lai, X. F.; Zhang, H.; Wang, W. Y. *Sci. Rep.* **2012**, *2*, 426.

(30) Ying, T. P.; Chen, X. L.; Wang, G.; Jin, S. F.; Lai, X. F.; Zhou, T. T.; Zhang, H.; Shen S. J.; Wang, W. Y. *J. Am. Chem. Soc.* **2013**, *135*, 2951-2954.

(31) Zhang, S. B.; Sun, Y. P.; Zhu, X. D.; Zhu, X. B.; Wang, B. S.; Li, G.; Lei, H. C.; Luo, X.; Yang, Z. R.; Song, W. H.; Dai, J. M. *Supercond. Sci. Technol.* **2009**, *22*, 015020.

(32) Mok, B. H.; Rao, S. M.; Ling, M. C.; Wang, K. J.; Ke, C. T.; Wu, P. M.; Chen, C. L.; Hsu, F. C.; Huang, T. W.; Luo, J. Y.; Yan, D. C.; Ye, K. W.; Wu, T. B.; Chang, A. M.; Wu, M. K. *Cryst. Growth Des.* **2009**, *9*, 3260-3264.

(33) Takeya, H.; Kasahara, S.; Hirata, K.; Mochiku, T.; Sato, A.; Takano, Y. *Physica C* **2010**, *470*, S497-S498.

(34) Hu, R. W.; Lei, H. C.; Abeykoon, M.; Bozin, E. S.; Billinge, S. J. L.; Warren, J. B.; Siegrist, T.; Petrovic, C. *Phys. Rev. B* **2011**, *83*, 224502.

(35) Chareev, D.; Osadchii, E.; Kuzmicheva, T.; Lin, J.-Y.; Kuzmichev, S.; Volkova, O.; Vasiliev, A. *CrystEngComm* **2013**, *15*, 1989.

(36) Böhmer, A. E.; Hardy, F.; Eilers, F.; Ernst, D.; Adelmann, P.; Schweiss, P.; Wolf, T.; Meingast, C. *Phys. Rev. B* **2013**, *87*, 180505(R).

(37) Rodríguez-Carvaja, J. *Physica B* **1993**, *192*, 55-69.

(38) Rotter, M.; Tegel, M.; Johrendt, D. *Phys. Rev. Lett.* **2008**, *101*, 107006.

(39) Zhou, T. T.; Chen, X. L.; Guo, J. G.; Jin, S. F.; Wang, G.; Lai, X. F.; Ying, T. P.; Zhang, H.; Shen, S. J.; Wang, S. C.; Zhu, K. X. *J. Phys.: Condens. Matter* **2013**, *25*, 275701.

(40) Nekrasov, I. A.; Sadovskii, M. V. *JETP Lett.* **2011**, *93*, 166-169.

(41) Richard, P.; Sato, T.; Nakayama, K.; Takahashi, T.; Ding, H. *Rep. Prog. Phys.* **2011**, *74*, 124512.

(42) Chen, F.; Xu, M.; Ge, Q. Q.; Zhang, Y.; Ye, Z. R.; Yang, L. X.; Jiang, J.; Xie, B. P.; Che, R. C.; Zhang, M.; Wang, A. F.; Chen, X. H.; Shen, D. W.; Hu, J. P.; Feng, D. L. *Phys. Rev. X* **2011**, *1*, 021020.

**Table 1.** Composition and $T_c^{onset}$ of SC $(Na_xK_y)Fe_zSe_2$ crystals

| Nominal composition | Composition by ICP-AES | $T_c^{onset}$ (K) |
|---|---|---|
| $Na_{0.75}Fe_{1.75}Se_2$ | $(Na_{0.16(1)}K_{0.70(1)})Fe_{1.72(3)}Se_{2.00(4)}$ | 29 |
| $Na_{0.75}Fe_{1.75}Se_2$ | $(Na_{0.18(1)}K_{0.66(1)})Fe_{1.68(3)}Se_{2.00(4)}$ | 32 |

**Figure 1.** The XRD pattern of crystal with composition of $(Na_{0.16}K_{0.70})Fe_{1.72}Se_2$. The reflections marked by asterisk are assigned to the SC phase. The inset shows the optical image of the as-prepared crystal embedded in the solution.

**Figure 2.** The SEM image and EDX results of freshly cleaved $(Na_{0.16}K_{0.70})Fe_{1.72}Se_2$ crystal. (a) Typical SEM image of $(Na_{0.16}K_{0.70})Fe_{1.72}Se_2$ crystal with SC stripes embedded in insulating background. (b) EDX results of Fe content vs (Na+K) content for SC stripes and background, respectively. (c) EDX results of K content vs Na content for SC stripes and background, respectively.

**Figure 3.** The magnetization of $(Na_{0.16}K_{0.70})Fe_{1.72}Se_2$ crystal as a function of temperature ranging from 5 to 80 K with $H$ parallel to $c$ axis. The inset shows the magnetic hysteresis measured at 5 K.

**Figure 4.** The temperature-dependent electrical resistance of $(Na_{0.16}K_{0.70})Fe_{1.72}Se_2$ crystal. The resistance for $K_{0.8}Fe_2Se_2$ [5] that was obtained by high-temperature route is also included for comparison. Humps are marked with arrows. Inset shows the enlargement of the low temperature region. $T_c^{onset}$ could be clearly observed at 29.5 K and $T_c^{zero}$ at 27.4 K. The resistance above $T_c^{onset}$ is fitted following the model proposed by Shoemaker and co-workers [24].

**Figure 5.** The Fermi surface and in-plane electronic band dispersion of $(Na_{0.16}K_{0.70})Fe_{1.72}Se_2$ recorded by ARPES with 21.2 eV photons from a helium discharge lamp along high symmetry lines. (a) Fermi surface displayed in the notation of 2 Fe/unit cell, with the corresponding Brillouin zone delimited by the dashed square. The intensity was integrated within $\pm 25$ meV with respect to $E_F$. (b)(c) ARPES intensity and MDCs plot recorded along "cut 1" in Fermi surface in (a), respectively. (d)(e) The same as (b)(c), but recorded along "cut 2" in Fermi

surface in (a). The blue dashed line in (b) and purple dashed line in (d) are guides to the eyes for the main band dispersions, whereas the blue dotted line in (c) and purple dotted line in (e) refer to the corresponding bands picked up from the MDCs, respectively.

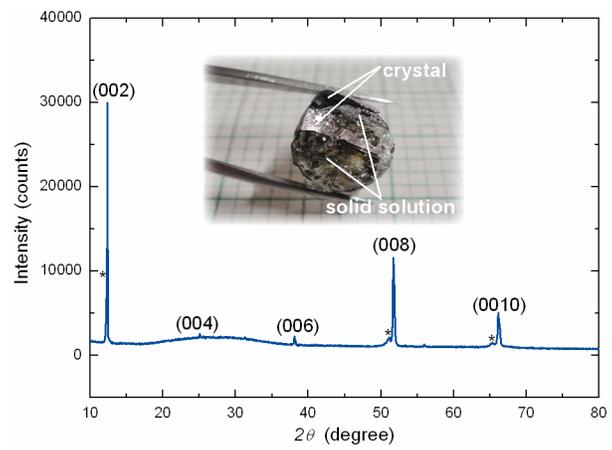

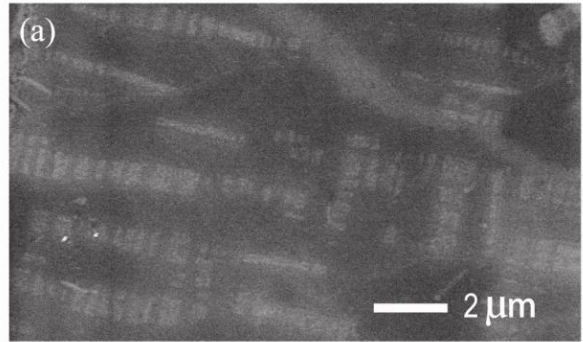

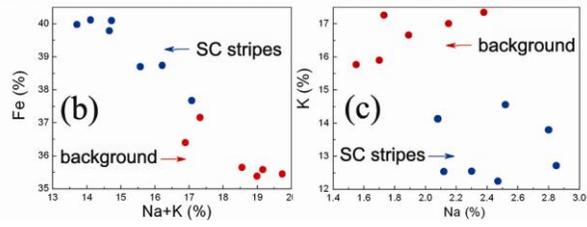

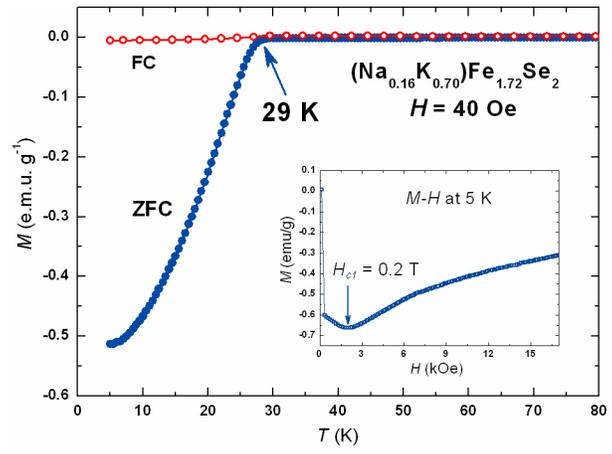

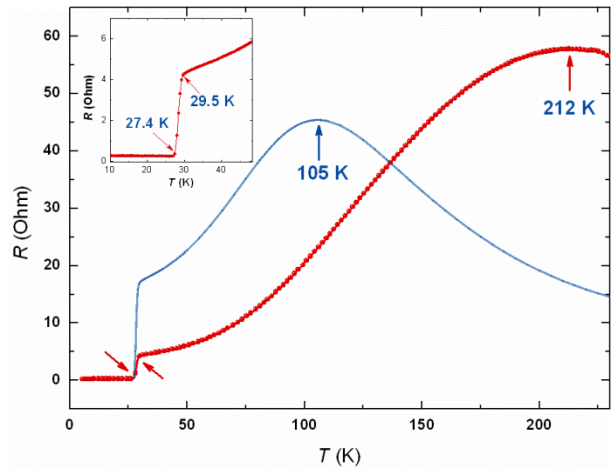

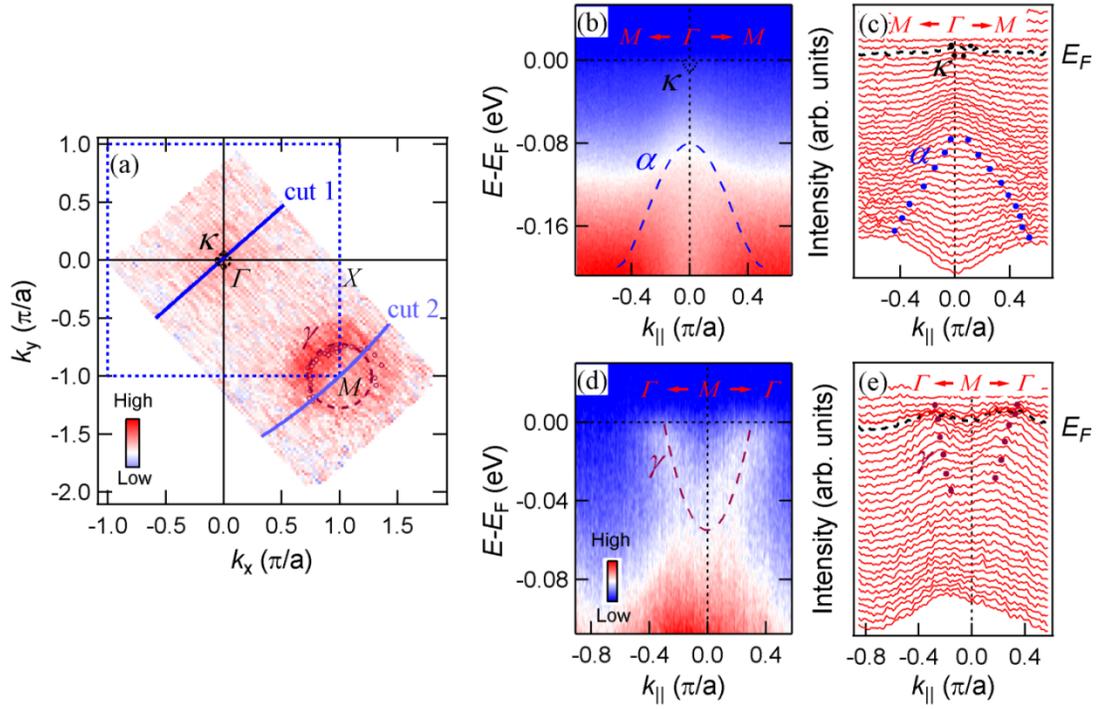